\documentclass[nofootinbib,prd,preprint,superscriptaddress]{revtex4}

\usepackage{axodraw}

\def\slashed#1{#1 \!\!\! /}

\begin{document}

\title{Iterative Structure Within The Five-Particle Two-Loop Amplitude}

\author{Freddy Cachazo}
\email{fcachazo@perimeterinstitute.ca}

\affiliation{Perimeter Institute for Theoretical Physics, Waterloo,
Ontario N2J 2W9, Canada}

\author{Marcus Spradlin}
\email{spradlin@ias.edu}

\affiliation{Institute for Advanced Study, Princeton NJ 08540, USA}
\affiliation{Michigan Center for Theoretical Physics, Ann Arbor MI 48104, USA}

\author{Anastasia Volovich}
\email{nastja@ias.edu}

\affiliation{Institute for Advanced Study, Princeton NJ 08540, USA}

\begin{abstract}
We find an unexpected iterative structure within the
two-loop five-gluon amplitude in
${\cal N} = 4$ supersymmetric Yang-Mills theory.
Specifically, we show that a subset of diagrams contributing to the
full amplitude, including a two-loop pentagon-box integral
with nontrivial dependence on five kinematical variables,
satisfies an iterative relation in terms of one-loop scalar box diagrams.
The implications of this result for the possible iterative
structure of the full two-loop amplitude are discussed.
\end{abstract}

\maketitle

\section{Introduction}

Scattering amplitudes in gauge theory exhibit simplicity that is
not manifest from Feynman diagrams~\cite{Witten:2003nn}.
This simplicity is even more striking
in the case of maximally supersymmetric ${\cal N}=4$ Yang-Mills
theory (MSYM).

One of the fascinating properties of MSYM is the possible presence of
iterative structures
relating amplitudes at different orders in perturbation theory.
In~\cite{Anastasiou:2003kj}
Anastasiou, Bern, Dixon and Kosower (ABDK) suggested
that the planar two-loop $n$-gluon
maximally helicity violating (MHV) amplitude obeys the iteration
\begin{equation}
\label{abdkintro}
M_n^{(2)}(\epsilon) =\frac{1}{2} \left(
M_n^{(1)}(\epsilon) \right)^2 + f^{(2)}(\epsilon)
M_n^{(1)}(2\epsilon) - \frac{\pi^4}{72} + {\cal O}(\epsilon),
\end{equation}
where $M_n^{(L)}(\epsilon) = A_n^{(L)}(\epsilon)/A_n^{(0)}$
is the ratio of the $L$-loop amplitude
(evaluated in $D = 4 - 2 \epsilon$ dimensions)
to the corresponding tree-level amplitude, and
\begin{equation}
f^{(2)}(\epsilon) \equiv
- \zeta(2) - \zeta(3) \epsilon - \zeta(4) \epsilon^2.
\end{equation}
This proposal was inspired by the
fact that the
collinear~\cite{Bern:1998sc,Kosower:1999xi,Kosower:1999rx,Anastasiou:2003kj}
and infrared singular~\cite{Magnea:1990zb,Catani:1998bh,Sterman:2002qn}
pieces of $M_n^{(2)}(\epsilon)$ are known to satisfy
an iterative relation of this form.
The statement that~(\ref{abdkintro}) actually holds for the full
two-loop amplitude $M_n^{(2)}(\epsilon)$ is the content of the ABDK
conjecture, which at present has been explicitly checked
only for the $n=4$ gluon amplitude~\cite{Anastasiou:2003kj}.
A similar iterative relation at three loops
was recently proven in~\cite{Bern:2005iz}, also for $n=4$.

The four-gluon amplitude is very special in the sense that it is a
nontrivial\footnote{There is a `trivial' overall
factor of $(s t)^{-\epsilon}$ in each term in
equation~(\ref{abdkintro}).}
function of a single dimensionless variable
$x= t/s$.
The $n=5$ case is much more complicated due to the presence of several
independent kinematical invariants. This is also the reason why one
would expect a richer, less rigid structure compared to the $n=4$
case.

Although the two-loop five-gluon amplitude is not known, a
conjecture for it has been given in~\cite{Bern:1997it}.
The conjecture for $M_5^{(2)}(\epsilon)$ has the
interesting property that contains two classes of terms,
\begin{equation}
M_5^{(2)}(\epsilon) = V_5^{(2)}(\epsilon) + W_5^{(2)}(\epsilon),
\end{equation}
where $V_5^{(2)}(\epsilon)$
is parity even and $W_5^{(2)}(\epsilon)$ is parity odd.
This is in contrast to the four-gluon amplitude
$M^{(2)}_4(\epsilon)$, which is wholly parity even.
In this paper we evaluate $V_5^{(2)}(\epsilon)$
explicitly through ${\cal O}(\epsilon^{-1})$ and
numerically\footnote{More specifically,
it is the relation~(\ref{csv}) that we check numerically
at~${\cal O}(\epsilon^0)$.
After this is done, it can of course be turned around
as in section III.B
and used to construct an explicit formula for
$V^{(2)}_5(\epsilon)$ through~${\cal O}(\epsilon^0)$.}
at ${\cal O}(\epsilon^0)$.
As one step in this calculation, we present an explicit
(through ${\cal O}(\epsilon^{-1})$)
formula for a two-loop pentagon-box
integral.  To our knowledge this is the first appearance
in the literature of a two-loop integral depending on five
kinematical invariants.
In performing this calculation we have benefited from
a program recently developed by
Czakon~\cite{Czakon:2005rk} which greatly facilitates the manipulation
and numerical evaluation of Mellin-Barnes integrals.

In order to check the relation~(\ref{abdkintro}) for $n=5$,
it is necessary to know the one-loop amplitude
$M_5^{(1)}(\epsilon)$
through ${\cal O}(\epsilon^2)$.  Through ${\cal O}(\epsilon^0)$,
it can be written as a linear combination of
one-loop scalar box integrals,
but starting at ${\cal O}(\epsilon)$ extra terms
appear
which can be expressed in terms of
pentagon integrals in
$D = 6 - 2 \epsilon$ dimensions~\cite{Bern:1992nf,Bern:1992em,Bern:1993kr}.
To our knowledge, the contribution of these extra pieces to
$M_5^{(1)}(\epsilon)$
has not been explicitly
computed.
We  therefore decompose
\begin{equation}
\label{wonedef}
M_5^{(1)}(\epsilon) = V_5^{(1)}(\epsilon) + W_5^{(1)}(\epsilon),
\qquad
W_5^{(1)}(\epsilon) = {\cal O}(\epsilon),
\end{equation}
where $V_5^{(1)}(\epsilon)$
consists of the one-loop scalar box terms and $W_5^{(1)}(\epsilon)$
contains the (currently unknown) extra pieces from the pentagon integrals.

In this paper we prove the remarkable fact that the $V_5^{(L)}(\epsilon)$
pieces alone
satisfy the ABDK relation~(\ref{abdkintro}), i.e.~we prove that
\begin{equation}
\label{csv}
V_5^{(2)}(\epsilon) =\frac{1}{2} \left(
V_5^{(1)}(\epsilon) \right)^2 + f^{(2)}(\epsilon)
V_5^{(1)}(2\epsilon) - \frac{\pi^4}{72} + {\cal O}(\epsilon).
\end{equation}

This paper is organized as follows:  In section II we
define the various integrals which are studied in this paper and
review the proposal~\cite{Bern:1997it} for the two-loop give-gluon
amplitude.
In section III we provide some details of the proof
of our main result~(\ref{csv}), postponing most technical details to the
appendix.
In section IV we use double-double unitarity cuts
to show that $M_5^{(2)}(\epsilon)$
must contain the term $W^{(2)}_5(\epsilon)$, and
we demonstrate the intriguing fact that this term has
very mild IR behaviour, $W^{(2)}_5(\epsilon) = {\cal O}(\epsilon^{-1})$.
In section V
we conclude with a discussion of our results and their
possible implications
for the iterative structure of the full two-loop amplitude.

\section{Five-Gluon Amplitudes at One and Two Loops}

In this section we present
formulas for the planar five-gluon amplitudes that enter into
the main result~(\ref{csv}).
Recall that for MHV amplitudes
it is very useful to normalize loop amplitudes
by the corresponding tree-level amplitudes.
Let us denote by $M_5^{(L)}(\epsilon)$ the normalized $L$-loop amplitude
$A_5^{(L)}(\epsilon)/A^{(0)}_5$.
We use the notation
\begin{equation}
s_i \equiv - (k_i + k_{i+1})^2.
\end{equation}

The one-loop five-gluon amplitude is known to
be~\cite{Bern:1992nf}
\begin{equation}
\label{oneloopfive}
M^{(1)}_5(\epsilon) = - \frac{1}{4} \sum_{\rm cyclic} \left[
s_3 s_4 I^{(1)}(\epsilon)
\right] + {\cal O}(\epsilon),
\end{equation}
where $I^{(1)}(\epsilon)$ is the one-loop scalar box integral
shown in Figure 1
and the sum is taken over the five cyclic permutations of
the external momenta.
The missing ${\cal O}(\epsilon)$
terms in~(\ref{oneloopfive}), which involve $D =  6 - 2 \epsilon$
pentagon integrals~\cite{Bern:1993kr},
are not explicitly known.  We therefore define
\begin{equation}
\label{vonedef}
V^{(1)}_5(\epsilon) \equiv
- \frac{1}{4} \sum_{\rm cyclic} \left[
s_3 s_4 I^{(1)}(\epsilon)
\right]
\end{equation}
with the understanding that $V^{(1)}_5(\epsilon)$
and $M^{(1)}_5(\epsilon)$ differ starting at
${\cal O}(\epsilon)$.  This difference is defined to be
$W^{(1)}_5(\epsilon) \equiv M^{(1)}_5(\epsilon)
- V^{(1)}_5(\epsilon)$.

\begin{figure}[t]
\begin{picture}(80,100)(-50,-110)
\put(-65,-64){\makebox(5,5){$I^{(1)} =$}}
\Boxc(0.000000,-62.000000)(30.000000,30.000000)
\ArrowLine(-15.000000,-77.000000)(-30.909903,-92.909903)
\put(-40.039029,-104.039029){\makebox(5.000000,5.000000){$k_1 + k_2$}}
\ArrowLine(-15.000000,-47.000000)(-30.909903,-31.090097)
\put(-40.039029,-24.960971){\makebox(5.000000,5.000000){$k_3$}}
\ArrowLine(15.000000,-47.000000)(30.909903,-31.090097)
\put(35.039029,-24.960971){\makebox(5.000000,5.000000){$k_4$}}
\ArrowLine(15.000000,-77.000000)(30.909903,-92.909903)
\put(35.039029,-104.039029){\makebox(5.000000,5.000000){$k_5$}}
\Line(-15.000000,-77.000000)(15.000000,-77.000000)
\end{picture}
\caption{The one-loop one-mass scalar box integral.  See appendix
A for details.}
\label{OneLoopIntegral}
\end{figure}
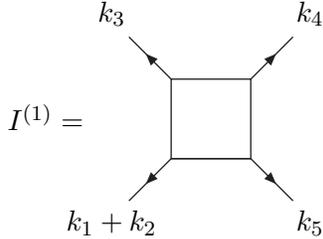

At two loops,
it has been conjectured~\cite{Bern:1997it} that\footnote{
Note that we have translated this expression into the more
modern normalization conventions
of~\cite{Anastasiou:2003kj,Bern:2005iz}.}
\begin{eqnarray}
\label{fifi}
M_5^{(2)}(\epsilon) &=& - \frac{1}{8}\sum_{\rm cyclic}
\left[
s_3 s_4^2 I^{(2)a}(\epsilon) + s_3^2 s_4 I^{(2) b}(\epsilon)
+ s_1 s_3 s_4 I^{(2) d}(\epsilon) + s_1 I^{(2) e}(\epsilon) \right],
\end{eqnarray}
where the various integrals are defined in Figure 2.
This formula for $M_5^{(2)}(\epsilon)$
can clearly be written as a sum of two kinds of terms,
\begin{eqnarray}
\label{vtwodef}
V_5^{(2)}(\epsilon ) &=&
- \frac{1}{8}\sum_{\rm cyclic}
\left[
s_3 s_4^2 I^{(2)a}(\epsilon) + s_3^2 s_4 I^{(2) b}(\epsilon)
+ s_1 s_3 s_4 I^{(2) d}(\epsilon)\right],
\label{wfive}
\\
\label{wproposal}
W_5^{(2)}(\epsilon )&=&- \frac{1}{8}
\sum_{\rm cyclic}
\left[
s_1 I^{(2) e}(\epsilon)
\right],
\end{eqnarray}
which are respectively parity even and parity odd.
The results of this paper provide very strong evidence
in support of the $V_5^{(2)}(\epsilon)$ part of this conjecture,
and in section 3 we show that the $W^{(2)}_5(\epsilon)$ term
must be present in $M^{(2)}_5(\epsilon)$ as well.

\begin{figure}[t]
\begin{picture}(220,110)(-110,-120)
\put(-65,-64){\makebox(5,5){$I^{(2)a} =$}}
\Boxc(0.000000,-62.000000)(30.000000,30.000000)
\ArrowLine(-15.000000,-77.000000)(-30.909903,-92.909903)
\put(-40.039029,-104.039029){\makebox(5.000000,5.000000){$k_1 + k_2$}}
\ArrowLine(-15.000000,-47.000000)(-30.909903,-31.090097)
\put(-40.039029,-24.960971){\makebox(5.000000,5.000000){$k_3$}}
\Boxc(30.000000,-62.000000)(30.000000,30.000000)
\ArrowLine(45.000000,-47.000000)(60.909903,-31.090097)
\put(65.039029,-24.960971){\makebox(5.000000,5.000000){$k_4$}}
\ArrowLine(45.000000,-77.000000)(60.909903,-92.909903)
\put(65.039029,-104.039029){\makebox(5.000000,5.000000){$k_5$}}
\end{picture}
\begin{picture}(220,110)(-110,-120)
\put(-65,-64){\makebox(5,5){$I^{(2)b} =$}}
\Boxc(0.000000,-62.000000)(30.000000,30.000000)
\ArrowLine(-15.000000,-77.000000)(-30.909903,-92.909903)
\put(-40.039029,-104.039029){\makebox(5.000000,5.000000){$k_5$}}
\ArrowLine(-15.000000,-47.000000)(-30.909903,-31.090097)
\put(-40.039029,-24.960971){\makebox(5.000000,5.000000){$k_1 + k_2$}}
\Boxc(30.000000,-62.000000)(30.000000,30.000000)
\ArrowLine(45.000000,-47.000000)(60.909903,-31.090097)
\put(65.039029,-24.960971){\makebox(5.000000,5.000000){$k_3$}}
\ArrowLine(45.000000,-77.000000)(60.909903,-92.909903)
\put(65.039029,-104.039029){\makebox(5.000000,5.000000){$k_4$}}
\end{picture}
\begin{picture}(130,95)(-20,-105)
\put(-65,-64){\makebox(5,5){$I^{(2)c} =$}}
\Boxc(0.000000,-62.000000)(30.000000,30.000000)
\ArrowLine(-15.000000,-77.000000)(-30.909903,-92.909903)
\put(-40.039029,-104.039029){\makebox(5.000000,5.000000){$k_{1}$}}
\ArrowLine(-15.000000,-47.000000)(-30.909903,-31.090097)
\put(-40.039029,-24.960971){\makebox(5.000000,5.000000){$k_{2}$}}
\Boxc(30.000000,-62.000000)(30.000000,30.000000)
\ArrowLine(45.000000,-47.000000)(60.909903,-31.090097)
\put(65.039029,-24.960971){\makebox(5.000000,5.000000){$k_{3}$}}
\ArrowLine(45.000000,-77.000000)(60.909903,-92.909903)
\put(65.039029,-104.039029){\makebox(5.000000,5.000000){$k_{4}$}}
\ArrowLine(15.000000,-77.000000)(15.000000,-99.500000)
\put(12.500000,-110.000000){\makebox(5.000000,5.000000){$k_{5}$}}
\end{picture}
\begin{picture}(130,95)(-110,-105)
\put(-70,-64){\makebox(5,5){$I^{(2)d} = (q - k_1)^2~\times $}}
\Boxc(0.000000,-62.000000)(30.000000,30.000000)
\ArrowLine(-15.000000,-77.000000)(-30.909903,-92.909903)
\put(-40.039029,-104.039029){\makebox(5.000000,5.000000){$k_{1}$}}
\ArrowLine(-15.000000,-47.000000)(-30.909903,-31.090097)
\put(-40.039029,-24.960971){\makebox(5.000000,5.000000){$k_{2}$}}
\ArrowLine(43.5317, -86.2705)(15,-77)
\ArrowLine(43.5317, -86.2705)(50.4846, -107.669)
\Line(43.5317, -86.2705)(61.1653, -62.0000)
\put(96.1653, -64.5000){\makebox(5.000000,5.000000){$k_{4}$}}
\ArrowLine(61.1653, -62.0000)(83.6653,-62)
\Line(61.1653, -62.0000)(43.5317, -37.7295)
\ArrowLine(43.5317, -37.7295)(50.4846,-16.331)
\Line(43.5317, -37.7295)(15,-47)
\put(50.3022, -117.302){\makebox(5.000000,5.000000){$k_{5}$}}
\put(50.3022,-11.6981){\makebox(5.000000,5.000000){$k_{3}$}}
\put(22.9542, -95.8666){\makebox(5.000000,5.000000){$q$}}
\end{picture}
\begin{picture}(230,120)(-155,-110)
\put(-123,-64){\makebox(5,5){$I^{(2)e} = (q - k_1)^2
\,{\rm tr}[\gamma_5\,({\slashed k}_1 +
{\slashed k}_2)\,{\slashed k}_3\,{\slashed q}_{\,}\,{\slashed k}_5 ]~\times $}}
\Boxc(0.000000,-62.000000)(30.000000,30.000000)
\ArrowLine(-15.000000,-77.000000)(-30.909903,-92.909903)
\put(-40.039029,-104.039029){\makebox(5.000000,5.000000){$k_{1}$}}
\ArrowLine(-15.000000,-47.000000)(-30.909903,-31.090097)
\put(-40.039029,-24.960971){\makebox(5.000000,5.000000){$k_{2}$}}
\ArrowLine(43.5317, -86.2705)(15,-77)
\ArrowLine(43.5317, -86.2705)(50.4846, -107.669)
\Line(43.5317, -86.2705)(61.1653, -62.0000)
\put(96.1653, -64.5000){\makebox(5.000000,5.000000){$k_{4}$}}
\ArrowLine(61.1653, -62.0000)(83.6653,-62)
\Line(61.1653, -62.0000)(43.5317, -37.7295)
\ArrowLine(43.5317, -37.7295)(50.4846,-16.331)
\Line(43.5317, -37.7295)(15,-47)
\put(50.3022, -117.302){\makebox(5.000000,5.000000){$k_{5}$}}
\put(50.3022,-11.6981){\makebox(5.000000,5.000000){$k_{3}$}}
\put(22.9542, -95.8666){\makebox(5.000000,5.000000){$q$}}
\end{picture}
\caption{Cast
of characters:
here we define the five two-loop integrals which appear in this paper.
All figures refer to the corresponding diagrams in a scalar field theory,
i.e., a collection of propagators integrated over the loop momenta.
The  integrals $I^{(2) d}$ and $I^{(2) e}$  are multiplied by
factors which depend on one of the loop momenta, as indicated
explicitly above.  These factors are meant to be included in the numerators
of the corresponding scalar integrals, transforming
them into tensor integrals. See appendix A for further details.}
\label{TwoLoopIntegrals}
\end{figure}
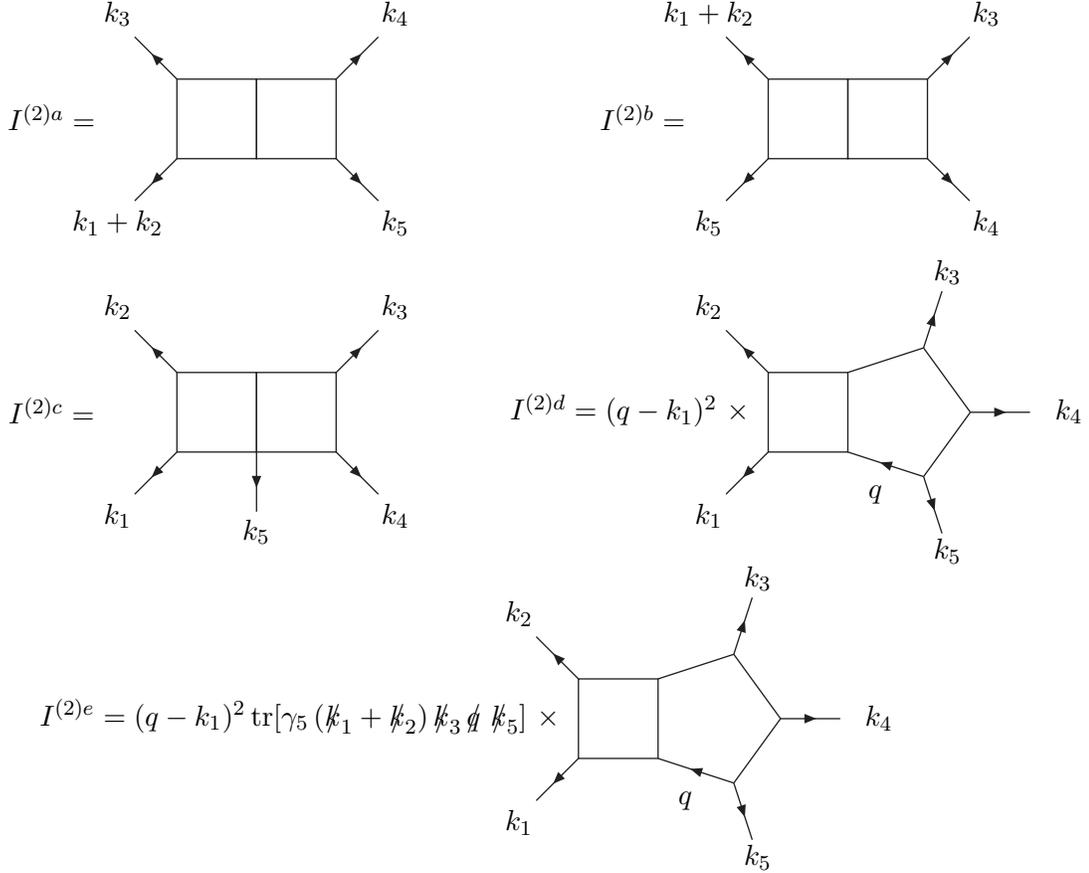

\section{Proof of the Main Result}

In this section we present our proof that $V_5^{(1)}(\epsilon)$
and $V_5^{(2)}(\epsilon)$,
defined in~(\ref{vonedef})
and~(\ref{vtwodef}) respectively, satisfy the ABDK relation~(\ref{csv})
\begin{equation}
V_5^{(2)}(\epsilon) =\frac{1}{2} \left(
V_5^{(1)}(\epsilon) \right)^2 + f^{(2)}(\epsilon)
V_5^{(1)}(2\epsilon) - \frac{\pi^4}{72} + {\cal O}(\epsilon).
\end{equation}
Our method is similar to that used in~\cite{Cachazo:2006mq}
to study the four-gluon amplitude $M_4^{(2)}(\epsilon)$ in that
we are able to check~(\ref{csv}) without
the need to fully evaluate any of the loop integrals in terms of
harmonic
polylogarithm functions\footnote{Although it is not
necessary for our proof, we nevertheless present
in appendix A
several explicit formulas in terms of polylogs, in particular
for
the integrals $I^{(2) c}$ and $I^{(2) d}$ which are new to the literature.}.
This is done by deriving the necessary
identities between various Mellin-Barnes integrals `under the integral
sign'.

We begin in
subsection III.A
by verifying~(\ref{csv})
explicitly through ${\cal O}(\epsilon^{-1})$.
That this works
is already highly nontrivial for two reasons.
First of all is the fact that the ansatz~(\ref{fifi}) for
$M_5^{(2)}(\epsilon)$ has not been completely proven, although our
results obviously provide strong evidence in its favor.
Second is the fact that, even if one assumes that the $M^{(L)}_5(\epsilon)$
satisfy~(\ref{abdkintro}), there is no obvious reason why this
should imply that the $V^{(L)}_5(\epsilon)$ satisfy~(\ref{csv}).
The difference between~(\ref{abdkintro})
and~(\ref{csv}) first shows up at ${\cal O}(\epsilon^{-1})$
because (as we show below) $W_5^{(2)}(\epsilon)$ starts contributing
to the left-hand side of~(\ref{csv}) at this order, and
$W_5^{(1)}(\epsilon)$ starts contributing to the right-hand side
of~(\ref{csv}) also at this order.

It seems a miracle that~(\ref{csv}) holds at ${\cal O}(\epsilon^{-1})$ even
though so many things might have spoiled it.
In subsection III.B we show that this miracle persists through
${\cal O}(\epsilon^0)$.
It seems clear that the ${\cal O}(\epsilon^0)$ terms in~(\ref{csv})
could, in principle, also be checked without fully evaluating the integrations
explicitly, but it would require a larger number of more complicated
identities.  For the purpose of this paper, which is to point out
an unexpected iterative structure within the full amplitude,
we are content
to perform the final step at ${\cal O}(\epsilon^0)$ numerically,
via a robust procedure described below.

\subsection{The ABDK Relation for $V_5^{(L)}(\epsilon)$
Through ${\cal O}(\epsilon^{-1})$}

Explicit formulas for the integrals $I^{(1)}(\epsilon)$
and $I^{(2)a}(\epsilon)$ have appeared
in the literature~\cite{Bern:1993kr,Smirnov:2000vy}.
We have evaluated the new integral $I^{(2)d}(\epsilon)$ explicitly through
${\cal O}(\epsilon^{-1})$.
The results for these integrals are summarized in appendix A, where
they are expressed in terms of several basic Mellin-Barnes-type
integrals which we call $A(a)$, $A_j(a)$ and $F(a,b)$.
The $A$ functions are defined by the integrals
\begin{eqnarray}
\label{adefone}
A(a) &=&
\int {dz \over 2 \pi i}\ a^z\, \Gamma^2(-z) \Gamma(z) \Gamma(1 + z),\\
A_1(a) &=& \int{dz \over 2 \pi i}\ a^z\, \Gamma^2(-z) \Gamma(z) \Gamma(1 + z)
( \psi(z) + \gamma),\\
A_2(a) &=& \int{dz \over 2 \pi i}\ a^z\, \Gamma^2(-z) \Gamma(z) \Gamma(1 + z)
( \psi(-z) + \gamma),\\
\label{adeffour}
A_3(a) &=& \int{dz \over 2 \pi i}\ a^z\, \Gamma^2(-z) \Gamma(z) \Gamma(1 + z)
{1 \over z},
\end{eqnarray}
where $\psi(x) =\frac{d}{dx}
 \ln \Gamma(x)$ and in each case $z$ is integrated
from $-i \infty$ to $+ i \infty$ along a contour which intersects the real
axis between $-1$ and $0$.
The function $F(a,b)$ is given by a more complicated double integral defined
in appendix A. Here we only need to use the fact that it
is symmetric in $a$ and $b$.
Explicit formulas for all of these functions in terms of harmonic
polylogarithms are given in appendix A.

If we plug the expressions given in appendix A
for the various integrals into~(\ref{csv}), we
find that the ${\cal O}(\epsilon^{-4})$
through ${\cal O}(\epsilon^{-2})$ pieces cancel easily,
leaving just
\begin{eqnarray}
\label{zero}
&& V^{(2)}_5(\epsilon) -\left[ \frac{1}{2} \left( V^{(1)}_5(\epsilon)\right)^2
+ f^{(2)}(\epsilon) V^{(1)}_5(2 \epsilon)\right]
\nonumber\\
&&\qquad=\frac{1}{\epsilon} \sum_{i=1}^5
\Bigg[
\frac{1}{4} A_1\left( \frac{s_i}{s_{i-2}} \right)
- \frac{1}{2} A_1\left( \frac{s_i}{s_{i+2}} \right)
- \frac{1}{4} A_2\left( \frac{s_i}{s_{i-2}} \right)
+ \frac{1}{2} A_2\left( \frac{s_i}{s_{i+2}} \right)
- \frac{1}{8} A_3 \left( \frac{s_i}{s_{i-2}} \right)
\nonumber\\
&&\qquad- \frac{1}{4} \ln \left( \frac{s_i}{s_{i+2}} \right)
A \left( \frac{s_i}{s_{i+2}} \right)
+ \frac{1}{8} \ln \left( \frac{s_i}{s_{i-2}} \right)
A \left( \frac{s_i}{s_{i-2}} \right)
- \frac{1}{24} \ln^3 \left(\frac{s_i}{s_{i-2}} \right)
\Bigg] + {\cal O}(\epsilon^0).
\end{eqnarray}
Interestingly, all of the $F(a,b)$ terms automatically drop out of
this expression due
to the symmetry $F(a,b) = F(b,a)$.  The remaining terms
on the last two lines
of~(\ref{zero}) can be seen to vanish with the help of the identities
\begin{eqnarray}
\label{identity}
A(a) + A(1/a) &=& - \frac{1}{2} \ln^2 (a) - \frac{\pi^2}{3},\\
A_1(a) - A_2(a) + \frac{1}{2} A_3(a)
&=& -\frac{1}{2} \ln(a) A(a),\\
A_1(1/a) - A_2(1/a) + \frac{1}{2} A_3(a)
&=&
- \frac{1}{2} \ln(a) A(a) - \frac{\pi^2}{3} \ln(a) - \frac{1}{3}
\ln^3(a),
\end{eqnarray}
which can be derived
by applying various tricks
directly to the definitions~(\ref{adefone})--(\ref{adeffour}).
This illustrates our point that it is possible to verify~(\ref{zero})
without the
need to fully evaluate loop integrals in terms
of harmonic polylogarithms.

An interesting fact, addressed in section V below,
is that the integral $I^{(2) c}$ does not seem to
appear in $V_5^{(2)}(\epsilon)$
(in the
chosen basis of integrals).
Suppose we make a more general ansatz
\begin{equation}
\label{ansatz}
- \frac{1}{8} \sum_{\rm cyclic} \left[
f_a s_3 s_4^2 I^{(2) a}(\epsilon)
 + f_b s_3^2 s_4 I^{(2) b}(\epsilon) + f_c s_1 s_2 s_3
I^{(2) c}(\epsilon) + f_d s_1 s_3 s_4 I^{(2) d}(\epsilon)
\right]
\end{equation}
and ask whether there exist numbers $f_a,\cdots,f_d$ such
that~(\ref{ansatz}) satisfies~(\ref{csv}).
At order ${\cal O}(\epsilon^{-4})$ and ${\cal O}(\epsilon^{-3})$
the only constraint is that the $f$'s should satisfy
\begin{equation}
f_a + f_b + \frac{9}{4} f_c + 3 f_d = 5.
\end{equation}
However, at ${\cal O}(\epsilon^{-2})$ the functional dependence of
the various integrals on the $s_i$ is much more complicated,
and it is
easy to see that the unique solution enabling~(\ref{ansatz}) to
satisfy~(\ref{csv}) even through ${\cal O}(\epsilon^{-2})$
is
\begin{equation}
f_a = f_b = f_d = 1, \qquad f_c = 0.
\end{equation}

\subsection{The ABDK Relation at ${\cal O}(\epsilon^0)$ and
The Finite Remainder}

Having demonstrated that $V^{(2)}(\epsilon)$ satisfies the ABDK
relation through ${\cal O}(\epsilon^{-1})$, let us now turn
our attention to the ${\cal O}(\epsilon^0)$ piece.
In principle one could continue as in the previous subsection by identifying
the basic integrals which emerge from the Mellin-Barnes representation
and then working out the necessary identities relating them to each other.
However, as mentioned above, it suffices for our purpose here to perform
this final step numerically.

A very efficient program for numerical evaluation of Mellin-Barnes
integrals has recently been developed by Czakon~\cite{Czakon:2005rk}
(see also~\cite{Anastasiou:2005cb}).
We repeatedly evaluated
each term in~(\ref{csv}) separately, for
randomly generated values of the kinematical variables $s_i$, and
always found that although each term in~(\ref{csv}) is generically
${\cal O}(1)$, the left-hand and the
sum of the terms on the right-hand side conspired to agree
with each other to within the precision of the numerical
integrations ($\sim 10^{-9}$).
This highly nontrivial conspiracy leaves us with no doubt that~(\ref{csv})
is correct.

A robust way to distill the information contained in~(\ref{csv})
at ${\cal O}(\epsilon^0)$ is to study the  so-called
finite remainder of $V^{(2)}(\epsilon)$.
The finite remainder of an arbitrary $L$-loop $n$-gluon amplitude
was defined in~\cite{Bern:2005iz}.
For the case at hand, we construct
the quantities
\begin{eqnarray}
\hat{I}^{(1)}(\epsilon)&\equiv&- \frac{1}{2 \epsilon^2} \sum_{i=1}^5
(s_i)^{-\epsilon},\\
\hat{I}^{(2)}(\epsilon)&\equiv& - \frac{1}{2} \left( \hat{I}^{(1)}(\epsilon)
\right)^2 + f^{(2)}(\epsilon) \hat{I}^{(1)}(2 \epsilon),
\end{eqnarray}
in terms of which the finite remainders for $V^{(L)}_5(\epsilon)$ are
defined by
\begin{eqnarray}
\label{fonedef}
F_5^{(1)}(\epsilon)&\equiv&V^{(1)}_5(\epsilon) - \hat{I}^{(1)}(\epsilon),\\
\label{finitedef}
F_5^{(2)}(\epsilon)&\equiv&V^{(2)}_5(\epsilon) -
\left[ \hat{I}^{(2)}(\epsilon) +
\hat{I}^{(1)}(\epsilon) V_5^{(1)}(\epsilon) \right].
\end{eqnarray}
Of course~\cite{Bern:2005iz} defined this quantity with the full
amplitudes $M^{(L)}_5(\epsilon)$ instead of
just the pieces $V^{(L)}_5(\epsilon)$ we are using here.
Nevertheless, the analysis of that paper proves that
if $V^{(2)}_5(\epsilon)$ satisfies~(\ref{csv}), then $F_5^{(2)}(\epsilon)$
must be given by
\begin{equation}
\label{ftwores}
F_5^{(2)}(\epsilon) =\frac{1}{2} \left(
F_5^{(1)}(\epsilon) \right)^2 + f^{(2)}(\epsilon)
F_5^{(1)}(2\epsilon) - \frac{\pi^4}{72} + {\cal O}(\epsilon).
\end{equation}
Of course it is simple to calculate $F_5^{(1)}(\epsilon)$
from the definitions~(\ref{vonedef}) and~(\ref{fonedef}),
\begin{equation}
F_5^{(1)}(\epsilon) = - \frac{1}{4} \sum_{i=1}^5
\ln\left(\frac{s_i}{s_{i+1}}\right)
\ln \left( \frac{s_{i-1}}{s_{i+2}}\right)
+ \frac{5 \pi^2}{8} + {\cal O}(\epsilon).
\end{equation}
Plugging this result into~(\ref{ftwores}) leads to the following
sharp prediction:
If~(\ref{csv}) holds at ${\cal O}(\epsilon^0)$, then
the quantity defined in~(\ref{finitedef}) must be equal to
\begin{equation}
\label{finiteresult}
F^{(2)}_5(\epsilon) =
\frac{89}{1152} \pi^4
+ \frac{11}{96} \pi^2 (X-Y) + \frac{1}{32} (X-Y)^2 + {\cal O}(\epsilon),
\end{equation}
where
we have defined
\begin{equation}
\label{xandy}
X = \sum_{i=1}^5 \ln^2\left(\frac{s_i}{s_{i+1}}\right), \qquad
Y = \sum_{i=1}^5 \ln^2\left(\frac{s_i}{s_{i+2}}\right).
\end{equation}

Having checked the ABDK relation~(\ref{csv}) as described above, the
formula~(\ref{finiteresult}) for the finite
remainder must also be true as a consequence.
Nevertheless we found it useful to also check~(\ref{finiteresult})
directly via a more robust numerical procedure as follows.
Supposing we didn't know~(\ref{finiteresult}) but wanted to check it
numerically,
we began with the ansatz that $F_5^{(2)}(0)$ should be
a quadratic polynomial in $\ln^2$ of kinematical invariants.
Now the right-hand side of~(\ref{finitedef})
is manifestly invariant under cyclic permutations $s_i \to s_{i+1}$,
so $F^{(2)}_5(\epsilon)$ must have this symmetry as well.
The objects $X$ and $Y$ defined in~(\ref{xandy}) are the only
independent quantities with the right structure satisfying this symmetry.
We therefore made the ansatz that $F_5^{(2)}(0)$ should be a quadratic
polynomial in $X$ and $Y$ with six coefficients to be determined,
\begin{equation}
\label{fansatz}
F^{(2)}_5(0) = a_1 \pi^4 + a_2 \pi^2 X
+ a_3 \pi^2 Y + a_4 X^2 + a_5 X Y + a_6 Y^2.
\end{equation}
The factors of $\pi^2$ were chosen conveniently so that all of the $a_i$
were expected to be rational numbers.
Indeed, we found that
the rational numbers $a_1 = 89/1152$, $a_2 = - a_3 = 11/96$ and $a_4
= a_5 = -  a_6/2 =1/32$ were
repeatedly obtained to a precision of $10^{-9}$ for various
randomly generated
values of the kinematical invariants, so we are quite confident
that the finite remainder $F^{(2)}_5(0)$ is indeed given
by~(\ref{finiteresult}).
The robustness of this numerical check comes from the fact that
we use numerical data to fix just six rational numbers.

\section{Consistency Checks on the $W_5^{(2)}(\epsilon)$ Piece of
$M^{(2)}_5(\epsilon)$}

In this section we study the structure of the proposed two-loop
five-gluon amplitude given in~\cite{Bern:1997it}. First we show that
$V^{(2)}_5(\epsilon)$ cannot possibly be equal to the full amplitude
$M^{(2)}_5(\epsilon)$ by itself.  This is done by computing a
double-double cut in a two-particle channel to see that the
$W^{(2)}_5(\epsilon)$ piece is needed. We then show that
$W_5^{(2)}(\epsilon)$ has very mild IR behaviour, diverging only as
${\cal O}(\epsilon^{-1})$. This fact makes the presence of
$W_5^{(2)}(\epsilon)$ consistent with the known
IR~\cite{Magnea:1990zb,Catani:1998bh,Sterman:2002qn} behaviour of
$M^{(2)}_5(\epsilon)$.

\subsection{Double-Double Cut}

Unitarity cuts provide very powerful constraints on scattering
amplitudes.
At one loop, amplitudes in supersymmetric gauge theories are
completely determined by
their cuts~\cite{Bern:1994zx,Bern:1994cg}.
This fact
is especially powerful if the basis of master integrals (after
reduction has been carried out) is known.
In particular, at
one loop in ${\cal N}=4$ SYM, the problem of computing any amplitude
is reduced to that of computing tree amplitudes by using quadruple
cuts~\cite{Britto:2004nc}.

At two loops, the basis of integrals is not known. In the case of
$n=4$ only double-box scalar integrals are needed to write the full
amplitude~\cite{Anastasiou:2003kj}.
This suggests that for $n=5$ the basis should
include the double boxes
$I^{(2) a}$, $I^{(2) b}$ and $I^{(2) c}$ shown in Figure 2.
In addition one might expect pentagon-boxes, i.e, a pentagon joined to a
box along one propagator. These are impossible for $n=4$ but for
$n=5$ the integral $I^{(2) d}$ shown in Figure 2 is natural.
One should also allow for the same classes of integrals with various tensor
structures in the numerator.

The proposal of~\cite{Bern:1997it} for $M_5^{(2)}(\epsilon)$ is a linear
combination of the integrals in Figure 2 with tensor
structures taking the general form
\begin{equation}
\label{tensor}
 A + B_\mu q^\mu + C_{\mu\nu} q^\mu q^\nu.
\end{equation}
Our goal in this subsection is to show that the pieces in
$V_5^{(2)}(\epsilon)$,
as defined in~(\ref{vtwodef}),
are not enough to be consistent with unitarity and that
the proposal of~\cite{Bern:1997it} is just right to give the correct cuts.

Consider the double two-particle unitarity cut in the $s_{1}$ channel.
Recall that unitarity cuts compute
discontinuities across branch cuts of the amplitude. In this case,
the double-double cut computes the discontinuity of the
discontinuity of the amplitude across the branch cut in the $s_{1}$
channel.
This is given by
(we use the notation $P_{ij\cdots} \equiv k_i + k_j + \cdots $)
\begin{eqnarray}
\label{double}
 C_{12}^{12} &=& \int \frac{d^D q}{(2 \pi)^D}\
\delta^{(+)}(q^2)\delta^{(+)}((q+P_{345})^2)A^{\rm
tree}(-(q+P_{345})^-,3^+,4^+,5^+,q^-)\nonumber\\
&&\times \int \frac{d^D \ell}{(2 \pi)^D}\
\delta^{(+)}(\ell^2)\delta^{(+)}((\ell-P_{12})^2) A^{\rm
tree}((-q)^+,\ell^-,(-\ell+P_{12})^-,(q+P_{345})^+)\nonumber\\
&&\qquad\qquad\qquad\qquad\times
A^{\rm tree}((-\ell)^+,1^-,2^-,(\ell-P_{12})^+).
\end{eqnarray}

The integral over $\ell$ is easy to recognize as the cut in the
$s_{1}$ channel of the one-loop amplitude
$A^{(1)}((-q)^+,1^-,2^-,(q+P_{345})^+)$. Therefore\footnote{
The strange prefactor in the last line converts 
from Feynman diagrams, which are normalized according to
$\int d^D p/(2 \pi)^D$, to the normalization~(\ref{boxdef}) for the
scalar box diagram.},
\begin{eqnarray}
\label{doublesim}
 C_{12}^{12} &=&
 \int \frac{d^D q}{(2 \pi)^D} \
\delta^{(+)}(q^2)\delta^{(+)}((q+P_{345})^2)A^{\rm tree}((-q)^+,1^-,2^-,
(q+P_{345})^+)\nonumber\\
&&\qquad\times A^{\rm
tree}(-(q+P_{345})^-,3^+,4^+,5^+,q^-)\nonumber\\
&&\qquad\times
\left[
i \frac{e^{-\epsilon \gamma}}{(4 \pi)^{2-\epsilon}}
\right]
s_{1}(q-k_1)^2 {\hbox{\lower 45pt\hbox{
\begin{picture}(80,100)(-50,-110)
\Boxc(0.000000,-62.000000)(30.000000,30.000000)
\ArrowLine(-15.000000,-77.000000)(-30.909903,-92.909903)
\put(-40.039029,-104.039029){\makebox(5.000000,5.000000){$(q + P_{345})^+$}}
\ArrowLine(-15.000000,-47.000000)(-30.909903,-31.090097)
\put(-40.039029,-24.960971){\makebox(5.000000,5.000000){$(-q)^+$}}
\ArrowLine(15.000000,-47.000000)(30.909903,-31.090097)
\put(35.039029,-24.960971){\makebox(5.000000,5.000000){$(k_1)^-$}}
\ArrowLine(15.000000,-77.000000)(30.909903,-92.909903)
\put(35.039029,-104.039029){\makebox(5.000000,5.000000){$(k_2)^-$}}
\Line(-15.000000,-77.000000)(15.000000,-77.000000)
\DashLine(0,-32)(0,-92){3}
\end{picture}
}}}
\end{eqnarray}

This expression can be further simplified by following the steps in
section 5 of~\cite{Bern:1994zx} where unitarity cuts of one-loop MHV
amplitudes are studied. The first step is to use that $q^2=0$ and
$(q+P_{345})^2=0$ to write
\begin{eqnarray}
\label{kewo}
&& A^{\rm
tree}((-q)^+,1^-,2^-,(q+P_{345})^+)A^{\rm
tree}(-(q+P_{345})^-,3^+,4^+,5^+,q^-) =\nonumber \\
&&\qquad\qquad\qquad {1\over 2}A^{\rm
tree}(1^-,2^-,3^+,4^+,5^+)\left[ -{{\rm tr}_{-}({\slashed
q}{\slashed k}_1({\slashed q}-{\slashed P}_{12}){\slashed k}_2)\over
(q-k_1)^2(q-k_1)^2} -{{\rm tr}_{-}({\slashed q}{\slashed
k}_5({\slashed q}-{\slashed P}_{12}){\slashed k}_2)\over
(q+k_5)^2(q-k_1)^2}
\right. \nonumber\\
&&\qquad\qquad\qquad\qquad\qquad\qquad\qquad\quad
\left. -{{\rm tr}_{-}({\slashed q}{\slashed k}_1({\slashed
q}-{\slashed P}_{12}){\slashed k}_3)\over (q-k_1)^2(q+P_{45})^2}
-{{\rm tr}_{-}({\slashed q}{\slashed k}_5({\slashed q}-{\slashed
P}_{12}){\slashed k}_3)\over (q+k_5)^2(q+P_{45})^2}\right],
\end{eqnarray}
where ${\rm tr}_-(\bullet) = \frac{1}{2}{\rm tr}((1-\gamma_5)\bullet)$.

Now we expand all terms inside the bracket in such a way that the
$q$-dependence is only in propagators or in ${\rm tr}(\gamma_5...)$ terms.
This immediately gives the result
\begin{eqnarray}
\label{almi}
 C_{12}^{12} &=& \frac{1}{2} \left[
i \frac{e^{-\epsilon \gamma}}{(4 \pi)^{2 - \epsilon}}\right]^2 A^{\rm
tree}(1^-,2^-,3^+,4^+,5^+)s_{1}\Big[(s_{5}s_{1}
-{\rm tr}(\gamma_5{\slashed q}{\slashed k}_5{\slashed k}_1{\slashed
k}_2))I^{(2) a}_{i \to i + 2} \nonumber \\
&&\qquad\qquad+  (s_{1}s_{2}
-{\rm tr}(\gamma_5{\slashed q}{\slashed k}_1{\slashed k}_2{\slashed
k}_3))I^{(2) b}_{i \to i + 3}
+(s_{3}s_{4} - {\rm tr}(\gamma_5{\slashed
q}{\slashed k}_3{\slashed k}_4{\slashed k}_5))
I^{(2) d}~\Big].
\end{eqnarray}
We use an abbreviated but hopefully transparent notation here:
On the first line, $I^{(2) a}_{i \to i + 2}$ means
the double-box integral $I^{(2) a}$ shown in Figure 2, but with
$k_i$ replaced by $k_{i+2}$ (so that, for example, the massive
leg has $k_3 + k_4$ instead of $k_1 + k_2$).
Moreover,
all three ${\rm tr}(\gamma_5 \cdots)$ terms are meant to appear in
the numerator of the corresponding integrand.
Finally, the three diagrams $I^{(2) a}$, $I^{(2) b}$ and
$I^{(2) d}$ appearing on the right-hand side of~(\ref{almi}) should
of course be double-double cut in the obvious way.

Note that except for the terms with ${\rm tr}(\gamma_5...)$, (\ref{almi})
would be consistent\footnote{Recall from equation~(1)
of~\cite{Anastasiou:2003kj} that $M_n^{(L)}(\epsilon)$
is normalized with a factor
of $\left[2 e^{-\epsilon \gamma}/(4 \pi)^{2 - \epsilon}\right]^{-L}$ relative
to Feynman diagrams.  Taking this into account turns the prefactor
in~(\ref{almi}) into $-\frac{1}{8}$, in agreement with~(\ref{vtwodef}).}
with $V^{(2)}_5(\epsilon)$. Luckily, the first two
${\rm tr}(\gamma_5...)$
terms give no contribution at all. To see this note that the whole
integral depends on only three independent momenta and by Lorentz
invariance the trace must give zero. For example, the first term
in~(\ref{almi}) only depends on $k_5$, $k_1$, and $k_2$.

Now it is clear why the last term does not vanish. The reason is
that the pentagon-box depends on four independent momenta.
This shows that the $W_5^{(2)}(\epsilon)$ in $M_5^{(2)}(\epsilon)$
has to be there in
order to be consistent with unitarity.

\subsection{Infrared Behaviour of $W_5^{(2)}(\epsilon)$}

Here we discuss the very special infrared behaviour that the parity
odd term $W_5^{(2)}(\epsilon)$ possesses.
The structure of IR
singularities as poles in $\epsilon$ was first studied at two loops
by Catani~\cite{Catani:1998bh}, who
showed that the coefficients of the
$1/\epsilon^4$, $1/\epsilon^3$ and $1/\epsilon^2$ terms in a two-loop
amplitude $M_n^{(2)}(\epsilon)$ are given
by\footnote{This is adapting Catani's
formula to the case of ${\cal N}=4$ SYM.}
\begin{equation}
\label{catani}
M_n^{(2)}(\epsilon) = \frac{1}{2} \left( M_n^{(1)}(\epsilon)
\right)^2 + f^{(2)}(\epsilon) M_n^{(1)}(2\epsilon) + {\cal O}(
\epsilon^{-1}).
\end{equation}
On the other hand, in this paper we have explicitly shown that~(\ref{csv})
\begin{equation}
\label{weshowed}
V_5^{(2)}(\epsilon) = \frac{1}{2} \left( M_n^{(1)}(\epsilon)
\right)^2 + f^{(2)}(\epsilon) M_n^{(1)}(2\epsilon) + {\cal O}(
\epsilon^{-1})
\end{equation}
(we have replaced $V_5^{(1)}(\epsilon)$ by $M_5^{(1)}(\epsilon)$
on the right-hand side since the difference $W_5^{(1)}(\epsilon)$
only starts contributing to this equation at ${\cal O}(\epsilon^{-1})$).
The only way to avoid a contradiction
between~(\ref{catani}) and~(\ref{weshowed}) is if the remaining
part $W_5^{(2)}(\epsilon) = M^{(2)}_5(\epsilon)
- V^{(2)}_5(\epsilon)$
is ${\cal O}(\epsilon^{-1})$.

By Lorentz invariance it must be true that
\begin{equation}
\label{newt}
W_5^{(2)}(\epsilon) = {\rm tr}(\gamma_5 {\slashed k}_1 {\slashed k}_2 {\slashed
k}_3 {\slashed k}_4) w_5^{(2)}(\epsilon)
\end{equation}
where $w_5^{(2)}(\epsilon)$ is a function of only the $s_{i}$ kinematical
invariants. To write $W_5^{(2)}(\epsilon)$ in the
form~(\ref{newt}) is straightforward using Feynman parameters where the
presence of the original $q^\mu$ only leaves a Feynman parameter
in the numerator.
It is easy to construct
a Mellin-Barnes representation for $w_5^{(2)}(\epsilon)$
(see appendix A.2).  For example, one can take the integral~(\ref{itwodmb})
with an additional factor of
\begin{equation}
\frac{\Gamma(-2 - 3 \epsilon - \nu - z_1)
\Gamma(-3 - 2 \epsilon - z_1 - z_3 - z_4 - z_5 - z_6)}
{\Gamma(-1 - 3 \epsilon - \nu - z_1)
\Gamma(-4 - 2 \epsilon - z_1 - z_3 - z_4 - z_5 - z_6)}
\end{equation}
in the integrand.
By explicit computation using the
program~\cite{Czakon:2005rk} we find the remarkable result that
\begin{equation}
\label{woow}
w_5^{(2)}(\epsilon) = {c_1\over \epsilon} + c_0 + {\cal
O}(\epsilon )
\end{equation}
where for example
\begin{eqnarray}
\label{epte}
c_1&\propto&\int \left[ \prod_{i=1}^4 \frac{dz_i}{2 \pi i}\Gamma(-z_i) \right]
\left( \frac{s_1}{s_2} \right)^{z_1}
\left( \frac{s_3}{s_2} \right)^{1 + z_2}
\left( \frac{s_4}{s_2} \right)^{1 + z_3}
\left( \frac{s_5}{s_2} \right)^{z_4}
\nonumber
\\&&\qquad\times
\Gamma(1 + z_1 + z_2) \Gamma(-1 - z_1 - z_2 - z_3) \Gamma(1 + z_1 + z_3)
\nonumber
\\&&\qquad\times
\Gamma(-1 - z_1 - z_2 - z_4) \Gamma(1 + z_2 + z_4)
\Gamma(2 + z_1 + z_2 + z_3 + z_4).
\end{eqnarray}
This is a rather surprising result, because one would expect
a generic two-loop tensor integral $q^\mu q^\nu$ to diverge as
$\epsilon^{-4}$.  This integral is special because
it contains the factor $q^\mu K_\mu$ in the numerator, where
the vector
$K_\mu = \epsilon_{\mu \nu \rho \sigma}
(k_1 + k_2)^\nu k_3^\rho k_5^\sigma$
has zero inner product with many of the external momenta.
One has to go all the way to ${\cal O}(\epsilon^{-1})$ before
a term can appear with sufficiently complicated structure to
survive being dotted with $K_\mu$.

\section{Conclusion and Discussion}

In this paper we have proven the striking result that $V_5^{(1)}(\epsilon)$
and $V_5^{(2)}(\epsilon)$ satisfy the ABDK relation~(\ref{csv}),
even though $V_5^{(1)}(\epsilon)$ is only part of the full one-loop amplitude
$M_5^{(1)}(\epsilon)$, and $V_5^{(2)}(\epsilon)$
is only part of the full two-loop
amplitude $M_5^{(2)}$.
Given this result,
it is easy to see that the full ABDK relation (\ref{abdkintro})
holds if and only if the `extra' pieces $W_5^{(L)}(\epsilon)\equiv
M^{(L)}_5(\epsilon) - V^{(L)}_5(\epsilon)$ are related by
\begin{equation}
\label{wsingular}
W^{(2)}_5(\epsilon) = V^{(1)}_5(\epsilon)
W_5^{(1)}(\epsilon) + {\cal O}(\epsilon) =
- \frac{5}{2 \epsilon^2}
(s_1 s_2 s_3 s_4 s_5)^{-\epsilon/5}\ W_5^{(1)}(\epsilon)
+ {\cal O}(\epsilon).
\end{equation}

In~\cite{Bern:1997it} it was conjectured that
\begin{equation}
\label{wfivetwo}
W_5^{(2)}(\epsilon) = - \frac{1}{8}
\sum_{\rm cyclic}
\left[
s_1 I^{(2) e}(\epsilon)\right].
\end{equation}
We have performed a couple of consistency checks on this conjecture.
First, we studied some unitarity cuts that explicitly show the
presence of the $W_5^{(2)}(\epsilon)$ term inside
$M_5^{(2)}(\epsilon)$. We also remarked that~(\ref{wfivetwo}) has a
very mild leading IR singularity ${\cal O}(\epsilon^{-1})$, which is
consistent with~(\ref{wsingular}) since $W_5^{(1)}(\epsilon)$ is
${\cal O}(\epsilon)$.

Finally, it is interesting to compare the formula for $M_5^{(2)}(\epsilon)$
in~(\ref{fifi})
with the results obtained by using the recent technique of octa-cuts
introduced in~\cite{Buchbinder:2005wp}.
The basic idea of the octa-cut technique is that at
two loops one can cut all propagators belonging to a box, i.e. perform a
quadruple cut~\cite{Britto:2004nc}
inside a double box or a pentagon-box and get from
the Jacobian a new propagator. The new propagator is just right so as to
produce a new effective one-loop box or pentagon integral. The remaining
one-loop object can then be cut four times to localize the final loop
integration momentum. In~\cite{Buchbinder:2005wp},
the coefficients of all double-boxes for the five-gluon amplitude
were computed. Perfect agreement is found with~(\ref{fifi}) for
the $I^{(2) a}$ and $I^{(2) b}$ integrals. However, there seem to be a
discrepancy for the $I^{(2) c}$ integral,
which is missing from~(\ref{fifi}) but should
have a nonzero coefficient according to~\cite{Buchbinder:2005wp}.

The solution to this puzzle is to realize that $I^{(2) d}$ secretly
contains an $I^{(2) c}$ integral. To see this, expand the numerator
$(q-k_1)^2=q^2-2k_1\cdot q$; the first term $q^2$ cancels a
propagator and produces precisely the $I^{(2) c}$
integral\footnote{Actually, a given $I^{(2) d}$ integral can be
decomposed in two different ways but we choose as the canonical
decomposition the one just described.}. We then find perfect
agreement with the octa-cut calculation of the $I^{(2) c}$
coefficient in~\cite{Buchbinder:2005wp}\footnote{Slightly more
subtle is to prove that~(\ref{fifi}) has the correct octa-cut in all
``channels". A more detailed analysis which is beyond the scope of
this paper shows that is the case.}.

This preliminary analysis shows that~(\ref{fifi}) satisfies several
non trivial constraints. As a future direction, it would be very
desirable to make a more detailed analysis of all unitarity cuts of
$M_5^{(2)}(\epsilon)$. It would also be interesting to compute
$W_5^{(1)}(\epsilon)$ in order to determine whether the
ansatz~(\ref{wfivetwo}) satisfies~(\ref{wsingular}).

\appendix

\section{Evaluation of Integrals}

This appendix
contains many of the technical details referred to in the
text.
The one- and two-loop integrals we need to study are shown
in Figures 1 and 2.
Our convention is that each loop momentum integral comes with a
normalization factor of
\begin{equation}
-i e^{\epsilon \gamma} \pi^{-D/2} \int d^D p,
\end{equation}
with $D = 4 - 2 \epsilon$.
This is the conventional normalization for studying iteration
relations and agrees with that used in writing~(\ref{abdkintro}),
although it differs
slightly from the normalization of amplitudes which should appear
in the actual $S$-matrix (see~\cite{Anastasiou:2003kj} for details).
As an illustration, the one-loop one-mass scalar box integral
of Figure 1 is defined to be
\begin{equation}
\label{boxdef}
I^{(1)}(\epsilon) \equiv
{\hbox{\lower 45pt\hbox{
\begin{picture}(80,100)(-50,-110)
\Boxc(0.000000,-62.000000)(30.000000,30.000000)
\ArrowLine(-15.000000,-77.000000)(-30.909903,-92.909903)
\put(-40.039029,-104.039029){\makebox(5.000000,5.000000){$k_1 + k_2$}}
\ArrowLine(-15.000000,-47.000000)(-30.909903,-31.090097)
\put(-40.039029,-24.960971){\makebox(5.000000,5.000000){$k_3$}}
\ArrowLine(15.000000,-47.000000)(30.909903,-31.090097)
\put(35.039029,-24.960971){\makebox(5.000000,5.000000){$k_4$}}
\ArrowLine(15.000000,-77.000000)(30.909903,-92.909903)
\put(35.039029,-104.039029){\makebox(5.000000,5.000000){$k_5$}}
\Line(-15.000000,-77.000000)(15.000000,-77.000000)
\end{picture}
}}}
= -i e^{\epsilon \gamma} \pi^{-D/2}
\int d^D p\
{1 \over p^2 (p - k_5)^2 (p + k_3 + k_4)^2 (p + k_4)^2}.
\end{equation}

We begin in subsection A.1 by recording our results for the various
integrals, before detailing the method by which they were obtained.
In subsection A.2 we present Mellin-Barnes representations
for the new integrals $I^{(2) c}$ and $I^{(2) d}$.
Finally in subsection A.3 we explain how we obtained the results of
A.1 by expressing everything in terms of a small number of
basic Mellin-Barnes-type integrals.

\subsection{Results}

In what follows we use the notation
\begin{equation}
s_i \equiv - (k_i + k_{i+1})^2, \qquad L_i \equiv \ln s_i
\end{equation}
and the functions
\begin{eqnarray}
\label{Adef}
A(a) &=& - H_{1,1}(1 - a) - H_{0,1}(1 - a) - \frac{\pi^2}{6},
\\
A_1(a) &=& -H_{1,1,1}(1-a) + H_{0,0,1}(1-a)
- {\pi^2\over 6} H_{1}(1-a),
\\
A_2(a) &=& H_{1,1,1}(1-a) + H_{1,0,1}(1-a) + H_{0,1,1}(1-a)
+ H_{0,0,1}(1-a) + \zeta(3),\\
A_3(a) &=& H_{1,1,1}(1-a) + H_{1,0,1}(1-a)
+ {\pi^2 \over 6} H_{1}(1-a) + 2 \zeta(3),
\end{eqnarray}
and
\begin{eqnarray}
\label{Fdef}
F(a,b) &=&
\frac{\pi^2}{6} H_{1}\left(1 - \frac{1}{a}\right)
- H_{0,0,1}\left(1 - \frac{1}{a}\right)
- H_{0,0,1}\left(\frac{1 - a - b}{1 - a}\right)
+ \frac{1}{2} H_{0,0,1}\left(\frac{1 - a - b}{(1 - a)(1 - b)}\right)
\nonumber\\
&&
+ H_{0,1,1}\left(\frac{1 - a - b}{1 - a}\right)
- \frac{1}{2} H_{0,1,1}\left(\frac{1 - a - b}{(1 - a)(1 - b)}\right)
+ H_{1,1,1}\left(1 - \frac{1}{a}\right)
\nonumber\\
&&
- H_{0,1}\left(\frac{1 - a - b}{1 - a}\right) \ln(1 - a)
+ H_{0,1}\left(\frac{1 - a - b}{(1 - a)(1 - b)}\right) \ln(1 - a)
- \frac{\pi^2}{6} \ln(a)
\nonumber\\
&&
- H_{0,1}\left(1 - \frac{1}{a}\right) \ln(a)
- H_{1,1}\left(1 - \frac{1}{a}\right) \ln(a)
+ \frac{1}{2} \ln^2(1 - a) \ln(1 - b)
\nonumber\\
&&
+ (a \leftrightarrow b).
\end{eqnarray}
We use here standard conventions for harmonic polylogarithm functions
(see for example the very useful program~\cite{Maitre:2005uu}).
It is very likely that the expression~(\ref{Fdef}) can be
simplified using various harmonic polylogarithm identities.
However, as we have emphasized, it is possible to explicitly
verify~(\ref{zero}) without knowing the precise formula~(\ref{Fdef}).
It turns out that as long as $F(a,b)$ is symmetric in $a$ and $b$,
which we have made manifest in~(\ref{Fdef}), then it automatically
drops out of~(\ref{zero}).

The one-loop one-mass scalar box integral $I^{(1)}$ was evaluated
to all orders in $\epsilon$ in~\cite{Bern:1993kr}.
In our notation it is given by
\begin{equation}
I^{(1)}(\epsilon) =\frac{ (s_3 s_4)^{-1-\epsilon}}
{(s_1)^{-\epsilon}}
 \left[ {2 \over \epsilon^2}
+ I^{(1)}_0 + \epsilon I^{(1)}_{+1} + {\cal O}(\epsilon^2)
\right]
\end{equation}
with
\begin{eqnarray}
I^{(1)}_0 &=& 2 A\left(  { s_{1} \over s_{3}} \right)
+ 2 A\left(  { s_{1} \over s_{4}} \right)
+ {\pi^2 \over 6},\nonumber
\\
I^{(1)}_{+1}&=&- 2 F\left(\frac{s_3}{s_1}, \frac{s_4}{s_1}\right)
- 2 \ln \left(\frac{s_1}{s_3}\right)  A_1\left(\frac{s_1}{s_3}\right)
- 2 \ln \left(\frac{s_1}{s_4}\right) A_1\left(\frac{s_1}{s_4}\right)
- \frac{14}{3} \zeta(3).
\end{eqnarray}

The two-loop integral $I^{(2) a}$ was evaluated
through ${\cal O}(\epsilon^0)$ in~\cite{Smirnov:2000vy}.
In our notation it is given by
\begin{equation}
I^{(2) a}(\epsilon) =
 {
 (s_{3})^{-1 - 2 \epsilon}
 (s_{4})^{-2 - 2 \epsilon}
\over
(s_{1})^{- 2\epsilon}
}
\left[ - {1 \over \epsilon^4}
+ \frac{I^{(2) a}_{-2}}{\epsilon^2}
+ \frac{I^{(2) a}_{-1}}{\epsilon}
+ {\cal O}(\epsilon^0)
\right],
\end{equation}
with
\begin{eqnarray}
I^{(2) a}_{-2} &=&
- A\left( {s_{1} \over s_{3}} \right)
- 3 A\left( {s_{1} \over s_{4}} \right)
-\frac{5 \pi^2}{12},\\
I^{(2) a}_{-1} &=&
2 F\left(\frac{s_3}{s_1},\frac{s_4}{s_1}\right)
+ 2 A_1\left(\frac{s_1}{s_3}\right)
+ 5 A_1\left(\frac{s_1}{s_4}\right)
- 2 A_2\left(\frac{s_1}{s_3}\right)
+ A_2\left(\frac{s_1}{s_4}\right)
- 3 A_3\left(\frac{s_1}{s_3}\right)
\nonumber\\
&&+ A_3 \left(\frac{s_1}{s_4}\right)
+ 2 \ln \left(\frac{s_1}{s_3}\right) A\left(\frac{s_1}{s_3}\right)
+ 4 \ln \left(\frac{s_1}{s_4}\right)  A\left(\frac{s_1}{s_4}\right)
+ \frac{25}{6} \zeta(3)
\end{eqnarray}

The two-loop integral $I^{(2) b}$ is clearly given simply by
\begin{equation}
I^{(2) b}(\epsilon)(s_1,s_3,s_4)
= I^{(2) a}(\epsilon)(s_1,s_4,s_3).
\end{equation}

The two-loop integral $I^{(2) c}$ has not appeared in
the literature.  We evaluate this integral\footnote{MS and AV
are grateful to R.~Roiban for collaboration
on early attempts to evaluate this integral.}
only through
${\cal O}(\epsilon^{-2})$, since that is sufficient to rule
out its appearance in the ABDK relation (see section III.A).  We find
\begin{equation}
I^{(2)c}(\epsilon) =
{
(s_{1} s_3)^{-1 - 2 \epsilon/3}
(s_{2})^{-1 - 2 \epsilon}
\over
(s_{4} s_5)^{-2 \epsilon/3}
}
\left[
  - {9 \over 4 \epsilon^4}
+ {I^{(2)c}_{-2} \over \epsilon^2}
+ {\cal O}(\epsilon^{-1})
\right],
\end{equation}
with
\begin{equation}
 I^{(2) c}_{-2} =
- A\left( {s_{4} \over s_{1}} \right)
- 2 A\left( {s_{4} \over s_{2} }\right)
- 2 A\left({ s_{5} \over s_{2} }\right)
- A\left({s_{5} \over s_{3}} \right)
+ {1 \over 2} \ln^2\left({s_{1} \over s_{4}} \right)
+ {1 \over 2} \ln^2\left( { s_{3} \over s_{5}} \right)
+ {5 \pi^2 \over 24}.
\end{equation}

Finally we have the integral $I^{(2) d}$, which also
has not appeared in the literature.  We find
\begin{equation}
I^{(2)d}(\epsilon) =
{(s_{2} s_5)^{-2 \epsilon/3}
(s_{3} s_4)^{-1- \epsilon/3}
\over s_1}
\left[  - {3 \over \epsilon^4} +
\frac{I^{(2)d}_{-2}}{\epsilon^2} +
\frac{I^{(2)d}_{-1}}{\epsilon}
+{\cal O}(\epsilon^0)\right],
\end{equation}
where
\begin{eqnarray}
\label{ipartd}
 I^{(2) d}_{-2} &=&
-3A\left({s_{1}\over s_{3}}\right)
- 2A\left({s_{4}\over s_{2}}\right)
- A\left({s_{3}\over s_{1}}\right)
- A\left({s_{4}\over s_{1}}\right)
- 3A\left({s_{1}\over s_{4}}\right)
- 2A\left({s_{3}\over s_{5}}\right)
- {\pi^2 \over 6}
\nonumber\\
&& -
\frac{4}{3} \ln^2\left(\frac{s_2}{s_5} \right)
+\frac{1}{6} \ln^2 \left(\frac{s_3}{s_4}\right)
- \frac{4}{3} (L_2 L_3 + L_4 L_5)
+ \frac{2}{3} (L_2 + L_3)(L_4 + L_5),
\end{eqnarray}
and
\begin{eqnarray}
I^{(2)d}_{-1} &=&
2 F\left(\frac{s_1}{s_3}, \frac{s_5}{s_3}\right)
+ 2 F\left(\frac{s_3}{s_1}, \frac{s_4}{s_1}\right)
+ 2 F\left(\frac{s_2}{s_4}, \frac{s_1}{s_4}\right)
\nonumber\\
&&
+ 6 A_1\left(\frac{s_1}{s_3}\right)
+ 2 A_1\left(\frac{s_3}{s_1}\right)
- A_1\left(\frac{s_1}{s_4}\right)
+ A_1\left(\frac{s_4}{s_1}\right)
\nonumber\\
&&
+ A_2\left(\frac{s_1}{s_3}\right)
- A_2\left(\frac{s_3}{s_1}\right)
+ 6 A_2\left(\frac{s_1}{s_4}\right)
-2 A_2\left(\frac{s_4}{s_1}\right)
- 2 A_2\left(\frac{s_4}{s_2}\right)
- 2A_2\left(\frac{s_3}{s_5}\right)
\nonumber\\
&&
+ 3 A_3\left(\frac{s_1}{s_3}\right)
+ A_3\left(\frac{s_3}{s_1}\right)
+ A_3\left(\frac{s_4}{s_1}\right)
+ 2 A_3\left(\frac{s_4}{s_2}\right)
+ 2 A_3\left(\frac{s_3}{s_5}\right)
- 4 A_3\left(\frac{s_5}{s_2}\right)
\nonumber\\
&&+
    (3 L_1 - 2 L_2 - 4 L_3 + L_4 + 2 L_5) A\left(\frac{s_1}{s_3}\right) -
    \frac{1}{3} (3 L_1 + 2 L_2 - 2 L_3 + L_4 - 4 L_5) A\left(\frac{s_3}{s_1}\right)
\nonumber\\
&&-
    \frac{1}{3} (3 L_1 - 4 L_2 + L_3 - 2 L_4 + 2 L_5) A\left(\frac{s_4}{s_1}\right) +
(2 L_2 + L_3 - L_4 - 2 L_5) A\left(\frac{s_1}{s_4}\right)
\nonumber\\
&& +
    \frac{2}{3} (L_2 + 2 L_3 + 2 L_4 - 5 L_5) A\left(\frac{s_4}{s_2}\right) -
    \frac{2}{3} (5 L_2 - 2 L_3 - 2 L_4 - L_5) A\left(\frac{s_3}{s_5}\right)
\nonumber\\
&& +
    \frac{2}{9} \pi^2 (L_2 + 2 L_3 + 2 L_4 - 5 L_5) +
    \frac{1}{27} (2 L_2 + L_3 + L_4 - 2 L_5)^3
\nonumber\\
&& -
    \frac{2}{3} (L_2 L_3^2 + L_2^2 L_4 + L_2 L_3 L_4 + L_3 L_5^2) +
    \frac{10}{9} L_2 (L_3 + L_4) L_5 +
    \frac{4}{9} L_3^2 L_5
\nonumber\\
&& +
    \frac{2}{9} (L_3 L_4 - L_4^2 - L_2^2) L_5 -
    \frac{2}{27} L_5^3 +
    14 \zeta(3)
\end{eqnarray}

\subsection{Mellin-Barnes Representations}

In order to evaluate these integrals we used the technology of
Mellin-Barnes (MB) integral representations (see~\cite{SmirnovBook}
for a thorough review).
It is straightforward to develop a MB representation
for all of the above integrals (see~\cite{Anastasiou:2005cb} for
a fairly general treatment).
For example, the new integrals $I^{(2) c}$ and $I^{(2) d}$ admit the
(highly non-unique)
MB representations
\begin{eqnarray}
I^{(2) c}(\epsilon) &=& - \frac{e^{2 \epsilon \gamma}}{s_5^{3 + 2
\epsilon}\Gamma(-2 \epsilon)}
\int \left[\prod_{i=1}^7 \frac{dz_i}{2 \pi i}
\Gamma(-z_i) \right]
\left( \frac{s_1}{s_5} \right)^{z_4}
\left( \frac{s_2}{s_5} \right)^{z_5}
\left( \frac{s_3}{s_5} \right)^{z_3 + z_7}
\left( \frac{s_4}{s_5} \right)^{z_6}
\nonumber\\&&\times
\frac{\Gamma(-1 - \epsilon - z_1 - z_2)}
{\Gamma(1 - z_1) \Gamma(-z_2) \Gamma(-1 - 3 \epsilon - z_3)} \Gamma(1 + z_2)
\Gamma(-1 - \epsilon - z_3)
\nonumber\\&&\times
\Gamma(1 + z_1 + z_3)
\Gamma(-1 - \epsilon + z_1 + z_2 - z_4 - z_5 - z_6)
\Gamma(1 + z_4 + z_6)
\nonumber\\&&\times
\Gamma(-z_2 + z_5 + z_6)
\Gamma(-2 - 2 \epsilon - z_3 - z_4 - z_6 - z_7)
\nonumber\\&&\times
\Gamma(1 - z_1 + z_4 + z_7)
\label{itwocmb}
\Gamma(3 + 2 \epsilon + z_3 + z_4 + z_5 + z_6 + z_7)
\end{eqnarray}
and
\begin{eqnarray}
I^{(2) d}(\epsilon) &=&
\lim_{\nu \to -1}
(-1)^\nu \frac{ e^{2 \epsilon \gamma} }{\Gamma(-2 \epsilon)}
\int \left[ \prod_{i=1}^7 \frac{dz_i}{2 \pi i}
\Gamma(-z_i)\right]
\left(\frac{s_1}{s_2}\right)^{z_1+z_5}
\left(\frac{s_3}{s_2}\right)^{z_6}
\left(\frac{s_4}{s_2}\right)^{z_4}
\left(\frac{s_5}{s_2}\right)^{z_7}
\nonumber\\
&&\times
\frac{\Gamma(-1 - \epsilon - z_1) \Gamma(1 + z_1 + z_2)
\Gamma(-1 - \epsilon - z_2 - z_3)}{\Gamma(-2 - 3 \epsilon - \nu - z_1)
\Gamma(1 - z_2) \Gamma(\nu - z_3)
\Gamma(3 + \epsilon + z_1 + z_2 + z_3)}
\nonumber
\\
&&\times
\Gamma(1 + z_3) \Gamma(2 + \epsilon + z_1 + z_2 + z_3)
\Gamma(1 - z_2 + z_4 + z_5)
\nonumber
\\&&\times \Gamma(-4 - 2 \epsilon - z_1 - z_3 - z_4 - z_5 - z_6)
\Gamma(3 + \epsilon + z_1 + z_2 + z_3 + z_5 + z_6)
\nonumber
\\&&\times\Gamma(-3 - 2 \epsilon - \nu - z_1 - z_5 - z_6 - z_7)
\Gamma(1 + z_6 + z_7)
\nonumber
\\&&\times
\label{itwodmb}
\Gamma(4 + 2 \epsilon + \nu + z_1 + z_4 + z_5 + z_6 + z_7).
\end{eqnarray}
A few comments are in order.  In each case the contours of integration
for the $z_i$ variables can be taken to be lines parallel to the imaginary
axis, as long as ${\rm Re}(z_i)$ are chosen such that the arguments
of all $\Gamma$ functions have a positive real part.
In the expression~(\ref{itwodmb}), the quantity
$\nu$ originates as the power of the factor $(q - k_1)^2$
in the denominator of the Feynman integral.
Since we want this factor to end up in the numerator, we want to take
$\nu \to -1$.  This limit must be defined by analytic continuation,
since it is not possible to satisfy all of the constraints on the
$z_i$ contours if one takes $\nu = -1$ at the start.

\subsection{Building Blocks}

A very efficient package for developing the $\epsilon$ series expansion
of a general Mellin-Barnes integral has recently been
developed by Czakon~\cite{Czakon:2005rk}.
Using this package, we can easily read off the values of the
desired integrals, order by order in $\epsilon$, in terms of some
basic building blocks.  Specifically, we find that
for all of the integrals considered in this paper,
only the following
two integral structures appear through order $\epsilon^{-2}$:
\begin{eqnarray}
\label{Aintdef}
A(a) &=&
\int {dz \over 2 \pi i}\ a^z\, \Gamma^2(-z) \Gamma(z) \Gamma(1 + z),\\
\label{Bintdef}
B(a) &=&
\int {dz \over 2 \pi i}\ a^{z + 1}\, \Gamma(-1 - z) \Gamma(-z)
\Gamma^2(1 + z).
\end{eqnarray}
The integration variable $z$ in each of these
expressions can be taken to run along
a line parallel to the imaginary axis from $-\frac{1}{4} - i \infty$
to $-\frac{1}{4} + i \infty$.
These integrals can be evaluated explicitly using standard
techniques.
For $a>0$ we find
$A(a)$ as written in~(\ref{Adef}), and
\begin{equation}
\label{Bdef}
B(a) = H_{0,1}(1-a) - \frac{\pi^2}{6}.
\end{equation}
We have chosen to express the argument of each dilogarithm
as $1 - a$ so that the branch cuts of the
integrals~(\ref{Aintdef}) and~(\ref{Bintdef})
for $a \in (-\infty,0)$ manifestly map on to the branch cuts
of the harmonic polylogarithms on the right-hand side.

Although we have written explicit formulas for $A(a)$ and $B(a)$
for the reader's benefit,
one of the points we want to emphasize in this paper
(following~\cite{Cachazo:2006mq}) is that
it is more efficient to check ABDK-type relations `under the integral
sign.'  For example, instead of evaluating the
integrals~(\ref{Aintdef}) and~(\ref{Bintdef}) explicitly, it is simpler
to notice that by taking $z \to - 1 -z$ inside the integral
we see immediately that
\begin{equation}
A(a) = B(1/a).
\end{equation}

This procedure of identifying the basic Mellin-Barnes building blocks
such as~(\ref{Aintdef}) and~(\ref{Bintdef}), and then deriving
various identities between them, can be carried out (with increasing
complication, of course) at each order
in $\epsilon$.  At ${\cal O}(\epsilon^{-1})$ a number of new integrals,
including double-integrals, start to appear.
Using various tricks `under the integral sign,' they can all be expressed
without too much difficulty in terms of the basic
integrals
\begin{eqnarray}
A_1(a) &=& \int{dz \over 2 \pi i}\ a^z\, \Gamma^2(-z) \Gamma(z) \Gamma(1 + z)
( \psi(z) + \gamma),\\
A_2(a) &=& \int{dz \over 2 \pi i}\ a^z\, \Gamma^2(-z) \Gamma(z) \Gamma(1 + z)
( \psi(-z) + \gamma),\\
A_3(a) &=& \int{dz \over 2 \pi i}\ a^z\, \Gamma^2(-z) \Gamma(z) \Gamma(1 + z)
{1 \over z},
\\
F(a,b) &=& \int {dz_1 \over 2 \pi i} {dz_2 \over 2 \pi i}
\ a^{1 + z_2} b^{1 + z_1}
\Gamma(-1 - z_1) \Gamma(-z_1) \Gamma(1 + z_1) \Gamma(-1 - z_2)
\nonumber\\
&&\qquad\qquad\qquad\times
\Gamma(-z_2) \Gamma(1 + z_2) \Gamma(2 + z_1 + z_2).
\end{eqnarray}
We have tabulated the explicit evaluations of these integrals
in subsection A.1, but we emphasize that it is possible to verify
the ABDK relation~(\ref{csv}) through ${\cal O}(\epsilon^{-1})$
without needing to evaluate these integrals,
due to the identities~(\ref{identity}).

\begin{acknowledgments}

We have benefited from valuable discussions and correspondence with
Z.~Bern, L.~Dixon, D.~Kosower and E.~Witten. In addition, FC is
grateful to E.~Buchbinder for useful discussions about some of the
generalized cuts of the ansatz for $M_5^{(2)}(\epsilon)$, and MS and AV are
grateful to R.~Roiban for collaboration on Mellin-Barnes integrals.
The research of FC at the Perimeter Institute is supported in part
by funds from NSERC of Canada and MEDT of Ontario. MS and AV
acknowledge support from the U.S.~Department of Energy under grant
number DE-FG02-90ER40542, and AV acknowledges support as the William
D.~Loughlin member of the IAS.

\end{acknowledgments}

\end{document}